\documentclass[runningheads]{svmult}

\usepackage{makeidx}   
\usepackage{graphicx}  
\usepackage{subeqnar}  
\usepackage{multicol}  
\usepackage{physprbb}  
\makeindex             

\newcommand{\greeksym}[1]{{\usefont{U}{psy}{m}{n}#1}}
\newcommand{\umu}{\mbox{\greeksym{m}}}

\begin{document}
\title*{Bolometric Light Curves of
Supernovae
}
\toctitle{Optical, Infrared, and Bolometric Light Curves 
\protect\newline of Type Ia Supernovae}
%
%
\titlerunning{Bolometric Light Curves of Supernovae}
%
\author{
Nicholas B. Suntzeff
}
\authorrunning{Nicholas Suntzeff}
\institute{
Cerro Tololo Inter-American Observatory,
National Optical Astronomy Observatories,
Casilla 603,
La Serena, Chile
}
\maketitle             

\begin{abstract}
 
The thermalized energy from the radioactive decays of $^{56}$Ni and
$^{57}$Ni and their daughter nuclides power the light curves of
supernovae near maximum light. The bolometric light curve gives us a
fundamental understanding of the energy evolution of a supernova
explosion and the amount of radioactive nuclides produced. In this
review, I will discuss the bolometric evolution of the Type IIp
supernovae SN1987A, and the general class of bolometric light curves
of Type Ia thermonuclear explosions. 

\end{abstract}

%

\section{SN1987A}

\begin{figure}[h]
\begin{center}
\includegraphics[angle=0,width=0.5\textwidth]{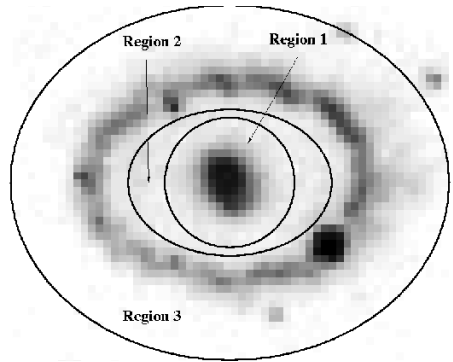}
\end{center}
\caption[nsuntzeff_fig01.eps]{An $HST$ image of region near the inner ring of
SN1987A. Region 1 isolates the expanding debris. Region 2 represents
the shocked region between the debris and the inner ring. Region 3
represents the inner ring}
\label{fig01}
\end{figure}

We have been monitoring the photometric properties of the Type IIp
SN1987A in the LMC for more than 15 years. Over the first 5 years the
optical photometry could be measured from the ground in the typical
1'' seeing at our observatories. In the near infrared bands of $JHK$
where the seeing is better and the crowding stars (which are typically
bluer than the SN) are less of a problem, we could monitor the SN for
up to 10 years. Once the supernova faded below the integrated
magnitudes of the inner ring which lies at about 1'' from the
supernova debris, $HST$ or ground-based AO imagery is needed to
isolate the debris and inner ring evolution. In Figure \ref{fig01}, I
show the structure of the inner ring region.

\begin{figure}[h]
\begin{center}
\includegraphics[angle=0,width=0.95\textwidth]{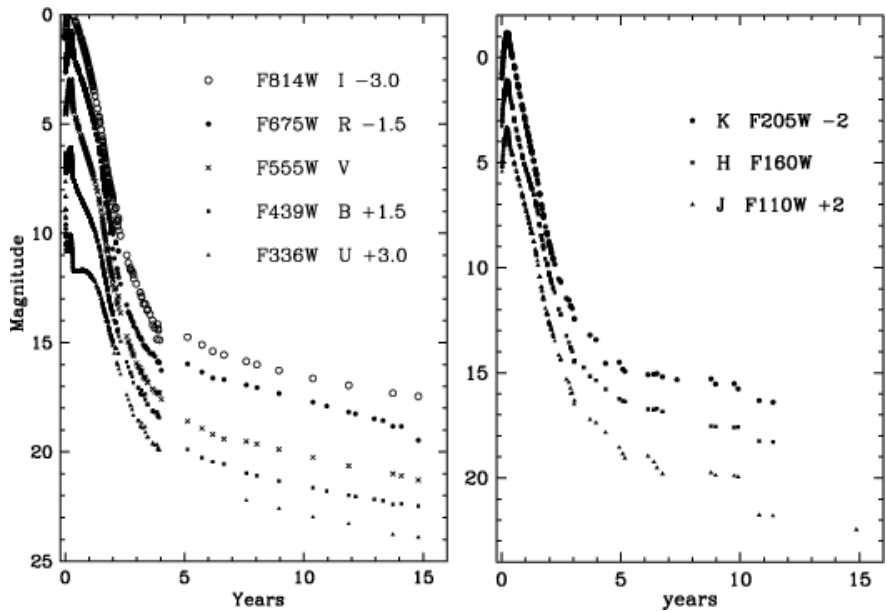}
\end{center}
\caption[nsuntzeff_fig02.eps]{Optical light curves (left panel) and
near-infrared light curves (right panel) for the debris (Region 1) in
SN1987A. The data have been shifted by the magnitudes listed in the
legend. The data are a combination of ground-based $UBVRIJHK$ and the
equivalent $HST$ flight system magnitudes.  }
\label{fig02}
\end{figure}

From our $HST$ images, we have measured the evolution of the
brightness of the three regions shown in Fig.~\ref{fig01} with respect
to an averaged background region outside of Region 3. For the earlier
ground-based data where we can only measure the light from all three
regions, we have subtracted off the extrapolated flux from Regions 2
and 3 to derive estimated fluxes for just the debris region. In
Fig.~\ref{fig02} we plot the optical light curve for 15 years of
evolution. With these corrections, the optical and $HK$ photometry
from the ground and $HST$ flight system equivalents merge well
together. The near-infrared photometry in $J/F110W$ shows a large
discontinuity presumably due to the radical differences in the two
filter systems. The photometry of non-stellar SEDs can show large
systematic errors due to the differences in the sensitivity functions
of the atmosphere/telescope/filter/detector system. In early
photometry of SN1987A, we noted difference of up to 0.4\,mag in $I$
photometry due to the differences in the facility $I$ filters
\cite{Sun_etal99}. In Table \ref{tab01}, I list the latest photometry
for SN1987A

\begin{table}
\caption{Optical Photometry of SN1987A as of May 2002}
\begin{center}
\renewcommand{\arraystretch}{1.0}
\setlength\tabcolsep{5pt}
\begin{tabular}{lccccc}
\hline\noalign{\smallskip}
& $U$ & $B$ & $V$ &$R$ &$I$ \\
\noalign{\smallskip}
\hline
\noalign{\smallskip}
debris & 20.91(32) & 20.88(12) & 21.28(07) & 20.96(22) & 20.46(04)\\
decay rate (mag y$^{-1}$) & 0.23  & 0.20  & 0.24  & 0.35  & 0.22 \\
\hline
\end{tabular}
\end{center}
\label{tab01}
\end{table}

\begin{table}
\caption{Near-Infrared Photometry of SN1987A}
\begin{center}
\renewcommand{\arraystretch}{1.0}
\setlength\tabcolsep{5pt}
\begin{tabular}{cccc}
\hline\noalign{\smallskip}
days since explosion& $J$ & $H$ & $K$ \\
\noalign{\smallskip}
\hline
\noalign{\smallskip}
4151 & 19.81(09) & 18.29(06) & 18.40(02)\\
5428 & 20.46(15) & ... & ... \\
\hline
\end{tabular}
\end{center}
\label{tab02}
\end{table}

The leveling off of the light curves after year 5 is due to two
causes: the longer decay times for the remaining radioactive nuclides
and the freeze-out of the cooling. By year five, the main radioactive
energy sources of $^{56}$Ni and its daughter nucleus $^{56}$Co with
e-folding times of 8.8d and 111.3d have decayed away. The nuclides
$^{57}$Co (the daughter nuclide of $^{57}$Ni) and $^{44}$Ti with
e-folding times of 390d and 87y are left as the main energy input,
along with the energy input from a possible (but as yet unseen)
pulsar. However, the light curves at this phase no longer represent
the prompt thermalization of the input radioactive energy. As shown in
\cite{Fra_Koz93,Fra_Koz01}, the recombination and cooling time scales
become larger than the expansion time scale, and the debris nebula is
not able to cool at the rate of radioactive energy input.

The ultraviolet, optical, and infrared photometry can be integrated to
the thermalized ``uvoir'' energy flux to derive the masses of the
synthesized radioactive nuclides. The early time data (less than
1000d) with prompt thermalization and cooling can be simply fit to
radioactive decay models as in
\cite{Sun_Bou90,Bou_etal91,Sun_etal92}. The late-time bolometric light
curves, when fit with models including the freeze-out, can be used to
measure the amount of $^{57}$Ni and $^{44}$Ti. Fransson and Kozma
(\cite{Fra_Koz01}) present the latest values for the amounts of these
nuclides where they found the following masses:
M($^{57}$Ni,$^{57}$Ni,$^{44}$Ti)=(0.069,0.003,0.0001)M$_\odot$.
Lundqvist et al.~(\cite{Lun_etal01}) find upper limits on the mass of
$^{44}$Ti consistent with these numbers based on the non-detection of
the 25m$\mu$ lines of [FeI] and [FeII] from ISO/SWS. They caution,
however, that if dust cooling is important, the limits on the mass of
$^{44}$Ti could be significantly higher. Only direct detection of the
1.157 MeV line of $^{44}$Ti will resolve the ambiguity in the models.

\begin{figure}[h]
\begin{center}
\includegraphics[angle=0,width=0.8\textwidth]{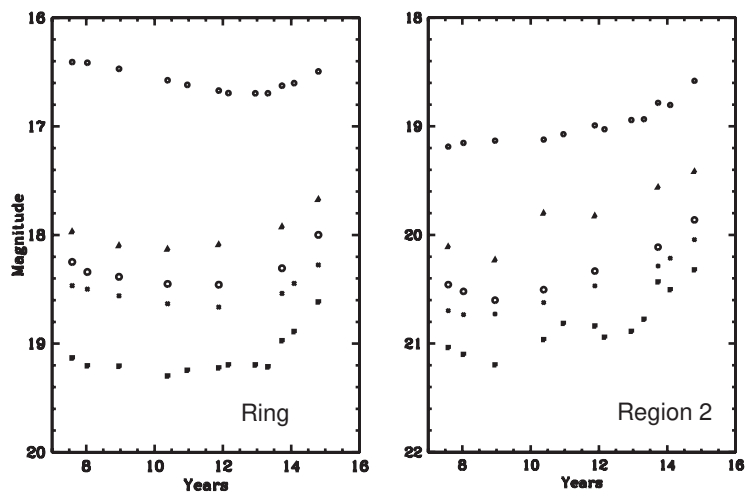}
\end{center}
\caption[nsuntzeff_fig03.eps]{$UBVRI$ optical light curves for the ring
Region 3 (left panel) and the inter-ring Region 2 (right panel). The
magnitudes from bottom to top are $BVIUR$. }
\label{fig03}
\end{figure}

In Fig.~\ref{fig03} I show the late-time photometric data for the ring
and the inter-ring region. The ring region suddenly began to brighten
at a rate of $\sim -0.24$ mag y$^{-1}$ in $UBVI$ and $\sim -0.12$ mag
y$^{-1}$ in $R$ since year 13. The brightening of the ring was
predicted in \cite{Lou_McC91,Che_Dwa95,Bor_etal97}. The ring was
expected to brighten by 7.5mag to about $V=11$ as the blast wave from
the debris struck the inner ring starting around year 16-20. However,
in year 10, Pun et al.~(\cite{Pun_etal97}) discovered a single spot
brightening, evidently due to an inward protrusion of the inner
ring. Our observations here show that the general rapid brightening of
the ring began later, around year 13.

The inter-ring region has accelerated its brightening since year 13
and is now brightening at a rate of $\sim -0.12$ mag y$^{-1}$ in
$UBVI$. The inter-ring light is presumably due to the the emission
from the reverse shock which is located at $\sim75$\% of the radius of
the inner boundary of the inner ring.  This emission was predicted by
\cite{Bor_etal97} and discovered by \cite{Son_etal98} with STIS data
from $HST$. Refined models are presented in
\cite{Mic_etal98a,Mic_etal98b}. The reverse shock formed in the
equatorial plane as neutral hydrogen atoms streaming from the debris
at 15000 km s$^{-1}$ hit the ionized region at the inner surface of
the ring. There is also diffuse light present in the STIS data which
may be due to excitation by non-thermal particles accelerated by the
shock.

\begin{figure}[h]
\begin{center}
\includegraphics[angle=0,width=0.7\textwidth]{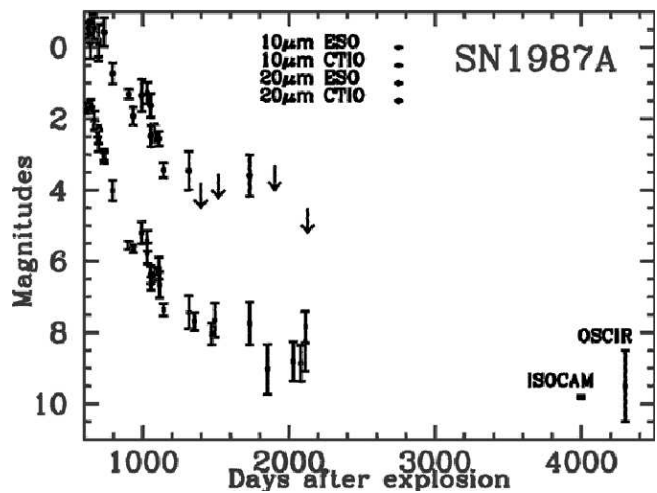}
\end{center}
\caption[nsuntzeff_fig04.eps]{Mid-infrared light curves of SN1987A}
\label{fig04}
\end{figure}

Significant dust formed in SN1987A starting around day 500, and by day
1000, the bolometric flux of the nebula was dominated by dust
radiation with a temperature of $T_{dust}\sim200$K and mid-infrared
line emission(\cite{Sun_etal92,Woo_etal93}). In Fig.~\ref{fig04} I
show the latest data on the mid-infrared evolution of SN1987A with the
ISOCAM detection (\cite{Fis_etal02}) and a detection from OSCIR on the
CTIO 4m telescope (\cite{Bou02}) at 10\umu{m}. Without 20\umu{m} detections
however, we can't estimate the dust temperature and cannot measure the
dust emission. If we assume the same bolometric corrections from day
1800, we find the bolometric flux of SN1987A is
${\rm{log}_{10}(L)\sim36.0}$ erg s$^{-1}$.

\section{The Bolometric Behavior of Type Ia Supernovae}

Because there is much less envelope mass above the radioactive energy
sources in Type Ia SNe, the bolometric properties of these supernovae
are more difficult to interpret because much less of the radioactive
energy is thermalized.  The physics of the thermalization is discussed
in \cite{Lei_Pin92,Lei00}.  The fraction of the radiation which is
thermalized must be transported to the surface and radiated away. If
the diffusion time for the energy is short compared to the dynamical
time, the energy appears as thermalized ``uvoir'' radiation. If the
diffusion time is long, the trapped radiation will be converted partly
into kinetic energy and not seen in the bolometric light curve. With a
longer diffusion time, the peak of the bolometric light curve is
shifted to later times when the input energy from the radioactive
decays is less. This leads to ``Arnett's Law'' \cite{Arn82} which
states that the bolometric luminosity at maximum light is equal to the
instantaneous energy released by the radioactive decay. This law was
(and is) an important tool to measure the amount of radioactive nickel
produced in the explosion.

Continuing the discussion in \cite{Lei_Pin92}, for the first 60 days
since explosion the thermalization time is short compared to the
dynamical time, and little radioactive energy is converted into
kinetic energy. Thus the integrated uvoir bolometric luminosity
represents the fraction of radioactive energy input which has been
thermalized.  By day 40 more than one half of gamma-ray radiation is
leaving the explosion unthermalized. Even at $B$ maximum light some 20
days after explosion, some 15\% of the gamma rays are
unthermalized. The observed uvoir bolometric light curve is then a
function of the rapidly decreasing optical depth to the gamma-rays
over the first 100 days and the decreasing energy input from the
radioactive nickel nuclides. The combination of these two processes
accelerates the light curve decay at a rate faster than the
radioactive decay of $^{56}$Co after maximum light.

The uvoir bolometric light curve interpretation is also complicated by
the efficiency of the trapping of the gamma rays and positrons in the
$^{56}$Co decay (\cite{Rui_Spr98,Mil_etal01,Lei_Sun03}). Milne et
al. (\cite{Mil_etal01}) show that the gamma-rays from the $^{56}$Co
decay carry 30 times more energy than the positrons, but also are more
penetrating and can escape more easily. As the nebula expands from
days 50 to 200, the energy deposition switches from the $^{56}$Co
gamma-ray Comptonization to the energy input from the positron
annihilations. Some of the positrons, however, can be transported out
of the nebula, depending on the magnetic fields structure. The
bolometric light curve at this phase then becomes dependent on the
details of the mixing of the $^{56}$Co in the nebula and the
properties of the magnetic fields.

\begin{figure}[h]
\begin{center}
\includegraphics[angle=0,width=1.0\textwidth]{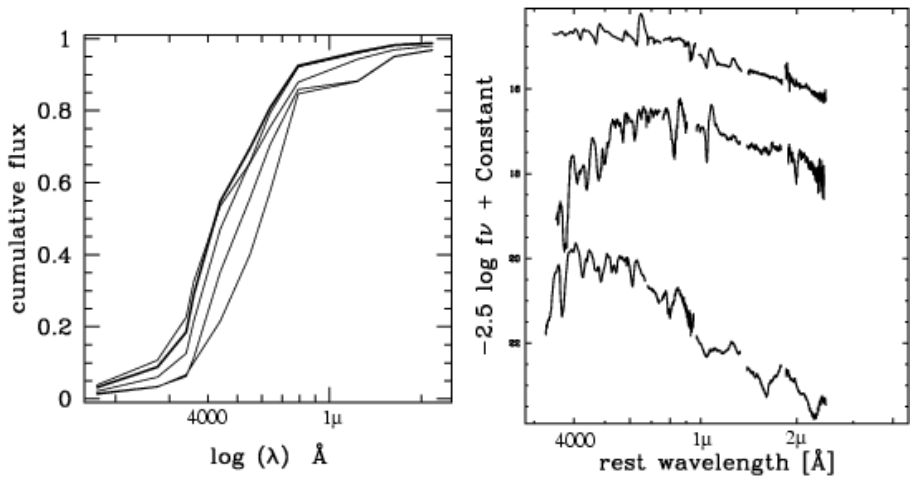}
\end{center}
\caption[nsuntzeff_fig05.eps]{Flux distributions of Type Ia SNe. The
left panel shows the cumulative flux distribution for SN1992A for days
--6,0,5,20,80 (from top to bottom at $\lambda=300$nm), with the
distribution for day 0 shown in the darker line. 80\% or more of the
total uvoir flux appears in the optical window of 300-1000nm. The
right panel shows the SEDs (top to bottom) for SN1999em (Type II), SN
1999ex (Ibc) and SN 1999ee (Type Ia) near maximum light (reference
\cite{Ham_etal02a}). Note that the broad flux peak of the SED appears
in the optical.}
\label{fig05}
\end{figure}

While the interpretation of the bolometric light curve requires
theoretical knowledge of the optical depths to gamma-rays and the
details about the mixing of $^{56}$Ni, the construction of the uvoir
bolometric light curve is rather simple due to the following
coincidence - - {\it most of the thermalized flux appears at optical
wavelengths.} In Fig.~\ref{fig05} I show the cumulative flux of
SN1992A at various phases during the first 100 days of evolution
(\cite{Sun96}). Also shown in the figure is a comparison of a Type Ia
SED with that of a Type Ic and II (\cite{Ham_etal02a,Ham_etal02b}). It
can be clearly seen that the broad peak of the flux distribution for
Type Ia SNe appears in the optical region. In fact, typically 80\% or
more of the uvoir flux appears in the optical from day --6 onwards.

Only a few papers have been published trying to estimate the
bolometric light curves (see \cite{Lei_Sun03} for a summary). In
\cite{Con_etal00,Vac_Lei96}, the optical broad-band magnitudes were
integrated to provide a magnitude that should be a close surrogate to
the uvoir bolometric magnitude. From the estimated bolometric light
curve, they found a range of more than a factor in 10 in the $^{56}$Ni
masses for a group of nearby Type Ia SNe.  In \cite{Cap_etal97}, a $V$
magnitude was used with a bolometric correction to study the gamma-ray
trapping in the late-time light curves, which also showed a
significant range in $^{56}$Ni masses.

In 1996, I used the small amount of data on supernovae which had space
ultraviolet spectra, optical $UBVRI$, and near-infrared $JHK$ data to
estimate accurate bolometric fluxes (\cite{Sun96}). I found that there
was a range in peak bolometric magnitudes (implying a range in nickel
masses) and that the bolometric light curves appear to have a small
secondary hump in the light curve around days 20-40 corresponding to
the secondary maximum in the $I$ band. Such a flux redistribution was
unexpected, and points to a rather sudden change in the opacity and
cooling in the nebula.

\begin{figure}[h]
\begin{center}
\includegraphics[angle=0,width=1.0\textwidth]{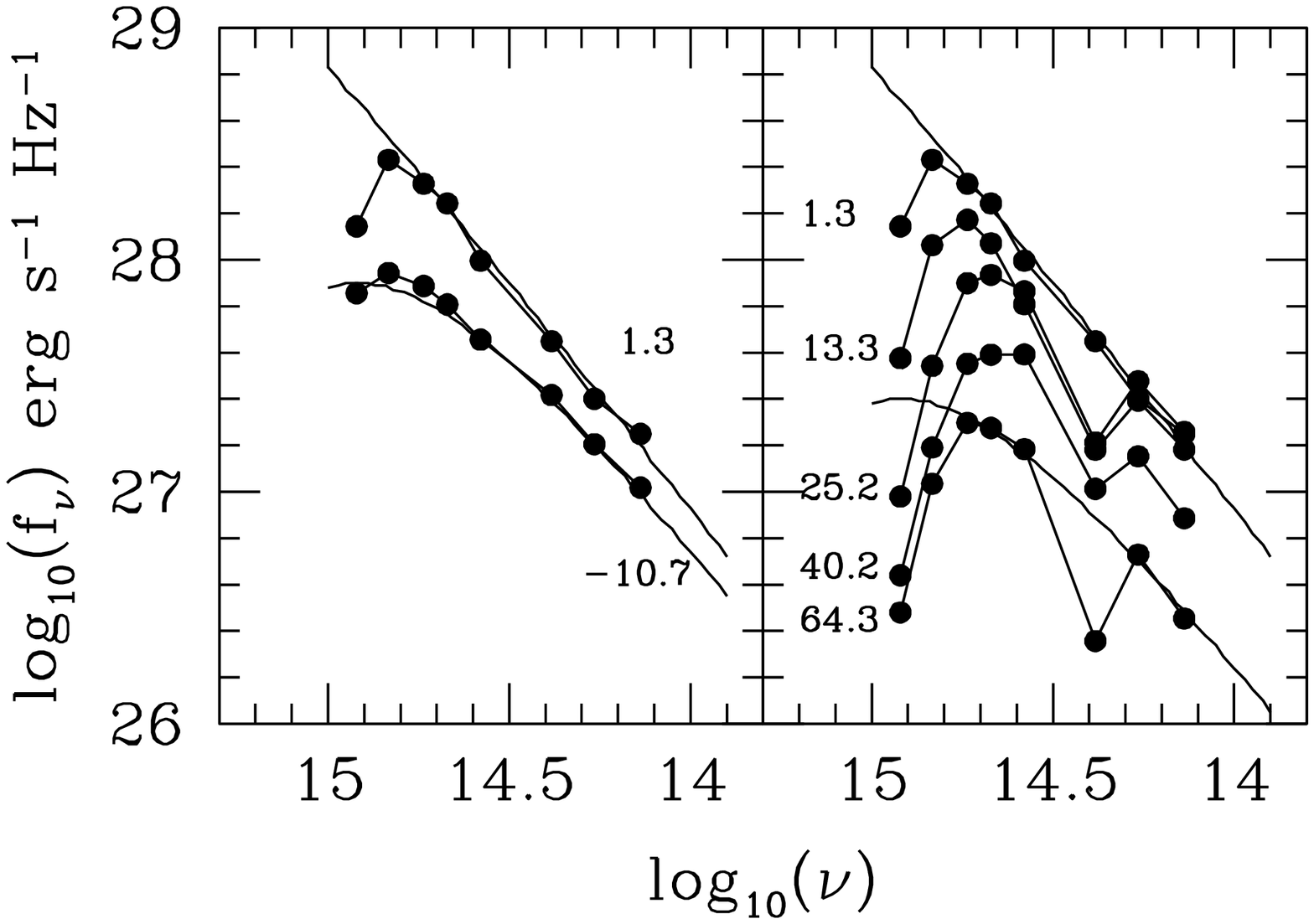}
\end{center}
\caption[nsuntzeff_fig06.eps]{SED evolution for the Type Ia SN2001el. The left
panels shows the SEDs for days (since $B_{max}$) --10.7 and +1.3. The
right panel shows the SEDs for days 1.3, 13.3, 25.2, 40.2, and 64.3. A
$T_{eff}=14,000$K blackbody is fit to the --10.7 and 64.3 data, and a
Rayleigh-Jeans law is fit to the data 1.3 data. The SED steepens to the
blue as the SN approaches maximum light and then reddens after
maximum. The flux deficit in $J$ appears right after maximum
light. Note that the $IHK$ data roughly follow the R-J law.
}
\label{fig06}
\end{figure}

The construction of the bolometric light curve requires the sum of the
space ultraviolet, optical, and near-infrared data. The falloff of the
flux beyond $H$ is such that the assumption of a R-J law adds only a
few percent to the total flux and a R-J extrapolation from $H$ or $K$
is entirely adequate. The extrapolation to the space ultraviolet,
however, is larger. Using 1992A data, I found that extrapolating the
$U$ flux point at 360nm to zero flux at 300nm was a reasonable
representation of the true space ultraviolet flux from day --6 onwards,
yielding agreement to 5\% in the uvoir flux between the full
integration and the integration using the ultraviolet extrapolation.
It is important to continue to observe nearby SNe with $HST$ in the
ultraviolet to understand the diversity of the ultraviolet flux, which
may be an indicator of the metallicity of the progenitor.

Our groups at CTIO and LCO have been following nearby Type Ia and Type
II in $UBVRIYJHK$ in part to study the bolometric properties of these
supernovae. The new filter $Y$ at 1\umu{m} (\cite{Hil_etal02}) has
been designed to fit in a wavelength band which is remarkably free of
telluric absorption and provides a flux point between the widely
separated $I$ and $J$ filters. Figure~\ref{fig06} (\cite{Kri_etal03})
shows the SED evolution of the nearby SN 2001el. This figure shows
that the SED is well fit by a black-body with $T_{eff}=14,000$K before
maximum light steepening to a R-J law by maximum light in
$BVRIJHK$. After maximum light, a flux deficit appears in $J$ and the
the bands $UBV$ drop away from a thermal distribution.

\begin{figure}[h]
\begin{center}
\includegraphics[angle=0,width=1.0\textwidth]{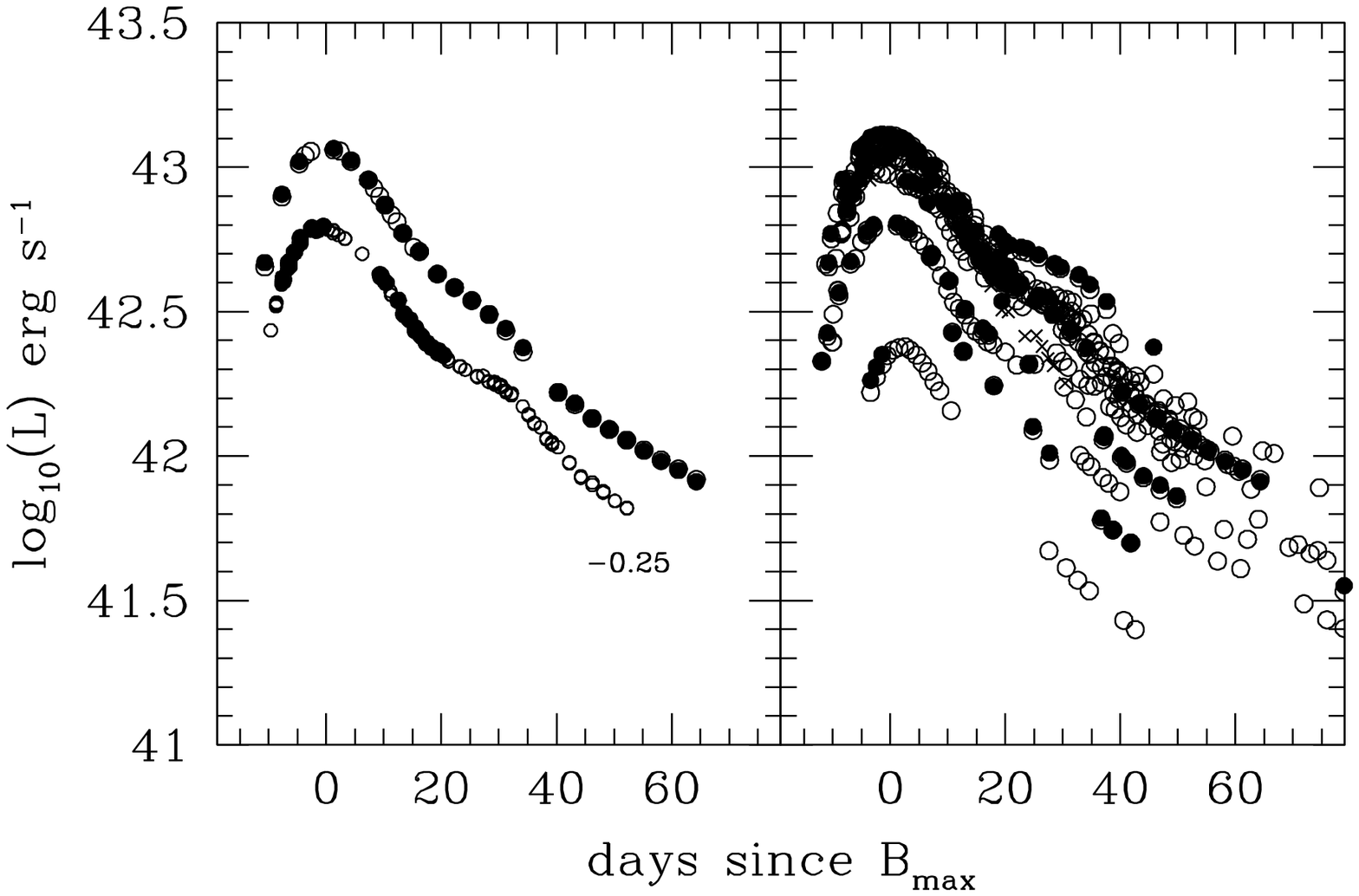}
\end{center}
\caption[nsuntzeff_fig07.eps] {``UVOIR'' bolometric light curves for Type Ia
SNe. These light curves include extrapolations to the space
ultraviolet and the extrapolation to the red using a R-J law. Closed
circles indicate an extrapolation from $K$ and open circles indicate
an extrapolation from $I$. The left panel shows the bolometric light
curves for SN2001el (upper curve, ${\rm\Delta}{m}_{15}=1.13$) and SN1999ee
(lower curve, ${\rm\Delta}{m}_{15}=0.94$). The SN1999ee data have been
shifted by 0.25dex for presentation purposes.  The right panel shows
the bolometric light curves for 16 Type Ia SNe.  }
\label{fig07}
\end{figure}

In Figure~\ref{fig07} I plot the uvoir bolometric light curves
calculated from the integration of the broad-band magnitudes,
including an extrapolation to the ultraviolet to account for the
unobserved space ultraviolet, and to the infrared using a R-J law
extending from $I$ or $K$. The absolute flux scale has been set by
using either SBF, Cepheid, TRGB, or PN distances for the nearby SNe,
or a Hubble law of $H_0=74$ km s$^{-1}$ Mpc$^{-1}$ for the more
distant SNe. The comparison of SN1999ee and SN2001el shows that the
location and size of the secondary bump is significantly different
despite the fact that the peak luminosities are very similar.

Also plotted in Fig.~\ref{fig07} are the uvoir bolometric light curves
for 16 Type Ia SNe. All have $UBVRI$ photometry and 10 have $JHK$
photometry. The sense of this figure is that most SNe have roughly the
same peak bolometric luminosity, but the size and placement of the
secondary hump is quite variable. For instance, the peculiar SN2000cx
has a rather normal bolometric light curve near maximum light, but
almost completely lacks a secondary hump at 30 days. This bolometric
behavior reflects the fact that SN2000cx had an anomalously weak
$I$-band secondary maximum. Evidently the peak bolometric magnitude
(and thus the $^{56}$Ni mass) is not strongly coupled to the opacity
and flux redistribution causing the secondary hump.

\begin{figure}[h]
\begin{center}
\includegraphics[angle=0,width=0.6\textwidth]{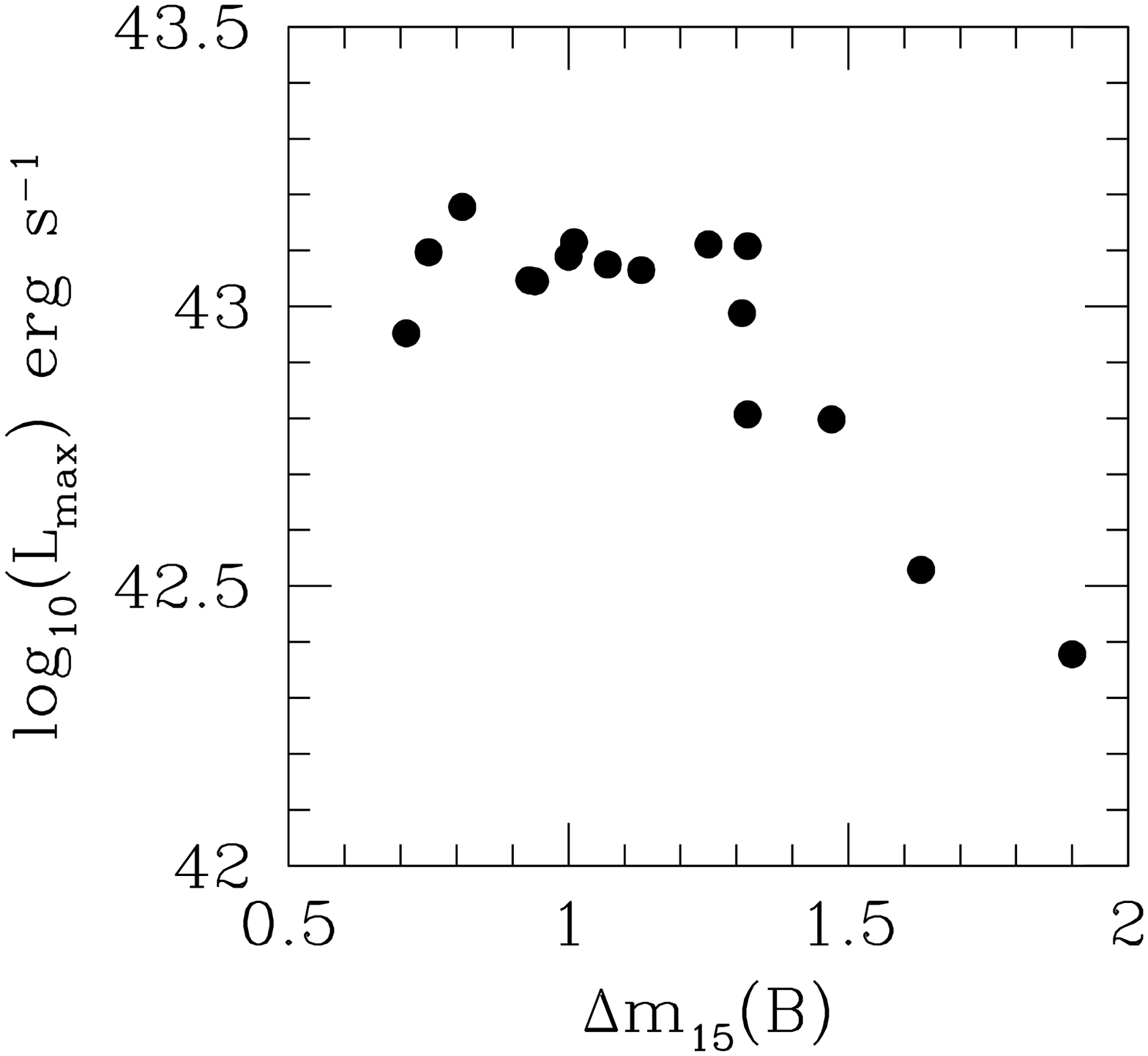}
\end{center}
\caption[nsuntzeff_fig08.eps] 
{
The peak uvoir bolometric luminosity plotted against the light curve
shape parameter of \cite{Phi93} for the SNe shown in Fig.~\ref{fig07}.
}
\label{fig08}
\end{figure}

Finally, in Fig.~\ref{fig08} I show the peak bolometric luminosity
versus the light curve shape parameter ${\rm\Delta}{m}_{15}$ for the sample
plotted in Fig.~\ref{fig07}. Rather than the roughly linear
relationship between the intrinsic luminosity in $B$ or $V$ and
${\rm\Delta}{m}_{15}$, we find that for much of the range of
${\rm\Delta}{m}_{15}$ there is little or no relationship between
${\rm{log}_{10}(L)}$ at peak and ${\rm\Delta}{m}_{15}$, at least for
${\rm\Delta}{m}_{15} < 1.3$. 

\smallskip

{\it Acknowledgments:} I would like to thank my collaborators:
M.~Phillips and M.~Hamuy (LCO); P.~Bouchet, P.~Candia, K.~Krisciunas,
R.~Schommer (deceased 12 December 2001), and C.~Smith, (CTIO);
B.~Leibundgut (ESO); R.~Kirshner, P.~Challis and the SInS
collaboration; and B.~Schmidt (MSSSO) and the High-Z Supernova Team
for their help in the collection of these data. This research was
supported in part by HST grants GO-07505.02A, GO-08177.6, and
GO08641.07A.

%

\end{document}